\documentclass[a4paper,10pt,twoside]{cpc-hepnp}

\usepackage{multicol}
\usepackage{graphicx}
\usepackage{booktabs}
\usepackage{amssymb,bm,mathrsfs,bbm,amscd}
\usepackage[tbtags]{amsmath}
\usepackage{lastpage}

\begin{document}

\fancyhead[c]{\small Chinese Physics C~~~Vol. xx, No. x (201x) xxxxxx}
\fancyfoot[C]{\small 010201-\thepage}


\title{Spectroscopy of $\Omega_{ccb}$ baryon in Hypercentral Quark Model}

\author{%
      Zalak Shah$^{1)}$\email{zalak.physics@gmail.com}
\quad Ajay Kumar Rai$^{2)}$\email{raiajayk@gmail.com}%
}
\maketitle

\address{%
Department of Applied Physics, Sardar Vallabhbhai National Institute of Technology, Surat, Gujarat, India-395007\\

}

\begin{abstract}
We extract the mass spectra of triply heavy bayon $\Omega_{ccb}$ using Hypercentral constituent quark model. The first order correction is also added to the  potential term of Hamiltonian. The radial and orbital excited state masses are also determined. Moreover, the Regge trajectories and magnetic moments are also given for this baryon. 
\end{abstract}

\begin{keyword}
Baryons, Potential Model, Regge trajectories
\end{keyword}

\begin{pacs}
12.39.pn, 12.40.Yx, 14.20.-c
\end{pacs}

\footnotetext[0]{\hspace*{-3mm}\raisebox{0.3ex}{$\scriptstyle\copyright$}2013
Chinese Physical Society and the Institute of High Energy Physics
of the Chinese Academy of Sciences and the Institute
of Modern Physics of the Chinese Academy of Sciences and IOP Publishing Ltd}%

\begin{multicols}{2}

\section{Introduction}
Last year, the affirmation of particles have been found in heavy hadron sector. LHCb has determined the five excited states of $\Omega_c$ baryon and the ground state of $\Xi_{cc}^{++}$ baryon \cite{lhcb, lhcb1}. Laterally, Belle-II also approved the excited states of $\Omega_c$ baryon \cite{belle}. Reviewing the heavy baryon sector, we can observe that the ground states of singly heavy baryons and two doubly heavy baryons are known experimentally till now \cite{olive}. And most of the excited state masses are well-known only for the singly heavy baryons in charm sector \cite{zcpc,Hua2016}. Now, we can expect such discovery of triply heavy baryons as well. The triply heavy baryon is a combination of three heavy quarks (c and/or b). These baryons are at the top most layer of SU(3) flavor symmetry. The spectroscopic properties of triply charm $\Omega_{ccc}$ and triply bottom $\Omega_{bbb}$ baryons have already been produced in our recent work \cite{zepja2}. In this article, we discuss, one of the triply heavy baryon which is $\Omega_{ccb}$. Many theoretical approaches have determined the masses of this baryon. They are, non-relativistic quark model \cite{Roberts2008}, Fadeev approach \cite{Bernotas2009}, Sum rules\cite{aliev2014,aliev2013,sun2015}, Bag model \cite{Hasan}, di-quark model \cite{kepjp}, Lattice QCD \cite{brown}, relativistic quark model \cite{Martynenko} and  variational cornell \cite{Flynn}.\\

The hypercentral constituent quark  model (hCQM) with color Coulomb plus linear potential has been used for systematic calculations of this baryon. Here, we have also incorporated the first order correction to the potential energy term of the Hamiltonian as it was in case of quarkonia \cite{koma}. The excited state mass spectra of heavy flavor baryons (singly, doubly and also some triply) are determined in our previous work using the same model  \cite{zepja, zepjc,zcpc,zbottom,zepjc2,zepja2}. Likewise, here, we determined the mass spectroscopy of $\Omega_{ccb}$ baryon which is a combination of two charm and one bottom quark. We also construct the Regge trajectories for these baryons in the ($n, M^2$) and ($J, M^2$) planes, where one can test several properties from the graph such as linearity, divergence, parallelism.  We also calculate the magnetic moment for the ground state mass. We calculate it for both $J^{P}= \frac{1}{2}^{+}$ and $J^{P}= \frac{3}{2}^{+}$ state. \\

This paper is organized as follows. A description of the hypercentral constituent quark model (hCQM) is given in Section 2. A systematic mass spectroscopy calculation has been performed in this model and we analyze and discuss our results in Section 3. We also plot the Regge trajectories and moreover, the magnetic moments are also determined. Finally, we draw conclusions in Section 4.

\section{The model}
The Hyper central Constitute Quark Model(hCQM) has successfully given the mass spectroscopy of the baryons in  heavy sector \cite{zepja, zepjc,zbottom,zepjc2,bhavin}.  We use the same methodology in this paper. The brief description of hCQM model is as follows:\\

The relevant degrees of freedom for the motion of heavy quarks are related by the Jacobi coordinates ($\vec{\rho}$ and $\vec{\lambda}$) are  \cite{Bijker,ginnani2015,8}
\begin{subequations}
\begin{equation}
\vec{\rho} = \dfrac{1}{\sqrt{2}}(\vec{r_{1}} - \vec{r_{2}})
\end{equation}
\begin{equation}
\vec{\lambda} =\dfrac{m_1\vec{r_1}+m_2\vec{r_2}-(m_1+m_2)\vec{r_3}}{\sqrt{m_1^2+m_2^2+(m_1+m_2)^2}}
\end{equation}
\end{subequations}
Here $m_i$ and $\vec{r_i}$ (i = 1, 2, 3) denote the mass and coordinates of the i-th constituent quark. Here, we only disscuss, $\Omega_{ccb}$ system, so that (i= 1,2) for $c$  quark and (i=3) for $b$ quark.

\noindent The Hamiltonian of system is defined as
\begin{equation}
H=\dfrac{P_{x}^{2}}{2m} +V(x)
\end{equation}
where $m$ is the reduced mass and $x$ is the six-dimensional radial hyper central coordinate of the three body system. The respective reduced masses are given by
\begin{subequations}
\begin{equation}
m_{\rho}=\dfrac{2 m_{1} m_{2}}{m_{1}+ m_{2}}
\end{equation}
\begin{equation}
 m_{\lambda}=\dfrac{2 m_{3} (m_{1}^2 + m_{2}^2+m_1m_2)}{(m_1+m_2)(m_{1}+ m_{2}+ m_{3})}
\end{equation}
\end{subequations}

\noindent In the center of mass frame ($R_{c.m.} = 0$), the kinetic energy operator can be written as
\begin{equation}
-\frac{\hbar^2}{2m}(\bigtriangleup_{\rho} + \bigtriangleup_{\lambda})= -\frac{\hbar^2}{2m}\left(\frac{\partial^2}{\partial x^2}+\frac{5}{x}\frac{\partial}{\partial x}+\frac{L^2(\Omega)}{x^2}\right)
\end{equation}
where $L^2(\Omega)$=$L^2(\Omega_{\rho},\Omega_{\lambda},\xi)$ is the quadratic Casimir operator of the six-dimensional rotational group O(6) and its eigenfunctions are the hyperspherical harmonics $Y_{[\gamma]l_{\rho}l_{\lambda}}(\Omega_{\rho},\Omega_{\lambda},\xi)$ satisfying the eigenvalue relation $L^2Y_{[\gamma]l_{\rho}l_{\lambda}}(\Omega_{\rho},\Omega_{\lambda},\xi)=-\gamma (\gamma +4) Y_{[\gamma]l_{\rho}l_{\lambda}}(\Omega_{\rho},\Omega_{\lambda},\xi)$. Here, $\gamma$ is the grand angular momentum quantum number. \\

The reduced six-dimensional hyperradial Schrodinger equation corresponding to the Eqn.~(2) of Hamiltonian can be written as
\begin{equation}\label{eq:6}
\left[\dfrac{-1}{2m}\dfrac{d^{2}}{d x^{2}} + \dfrac{\frac{15}{4}+ \gamma(\gamma+4)}{2mx^{2}}+ V(x)\right]\phi_{ \gamma}(x)= E\phi_{\gamma}(x).
\end{equation}

\noindent The hypercentral potential V(x) as the color Coulomb plus linear potential with first order correction as well as spin interaction is defined by \cite{koma,devlani,rai,12}
\begin{equation}
V(x) =  V^{0}(x) + \left(\frac{1}{m_{\rho}}+ \frac{1}{m_{\lambda}}\right) V^{(1)}(x)+V_{SD}(x) 
\end{equation}
\begin{equation}
V^{(0)}(x)= \frac{\tau}{x}+ \beta x ~~~~\&~~~~~ 
V^{(1)}(x)= - C_{F}C_{A} \frac{\alpha_{s}^{2}}{4 x^{2}}
\end{equation}

\begin{eqnarray}
V_{SD}(x)= V_{SS}(x)(\vec{S_{\rho}}.\vec{S_\lambda})
+ V_{\gamma S}(x) (\vec{\gamma} \cdot \vec{S})&&  \nonumber \\ + V_{T} (x)
\left[ S^2-\dfrac{3(\vec{S }\cdot \vec{x})(\vec{S} \cdot \vec{x})}{x^{2}} \right]
\end{eqnarray}
Here, the hyper-Coulomb strength $\tau = -\frac{2}{3} \alpha_{s}$ where $\alpha_{s}$ corresponds to the strong running coupling constant; $\frac{2}{3}$ is the color factor for baryons, and $\beta$ corresponds to the string tension for baryons.$C_{F}$ and $C_{A}$ are the Casimir charges of the fundamental and adjoint representation with values $\frac{2}{3}$ and 3. The spin-dependent part of Eqn.~(8), $V_{SD}(x)$, contains three types of  interaction terms, including the spin-spin term $V_{SS} (x)$, the spin-orbit term $V_{\gamma S}(x)$ and a tensor term $V_{T}(x)$ \cite{zepja}. We numerically solve the six-dimensional Schrodinger equation using Mathematica notebook \cite{lucha}.  The values of quark masses are $m_c$=1.275 GeV and $m_b$=4.67 GeV in calculation.  We fix the ground state mass and after that we calculate radial and orbtal excited state masses.\\
 
\begin{center}
\tabcaption{\label{tab:1} Ground state masses of $\Omega_{ccb}$ baryon (in GeV)}
\begin{tabular*}{80mm}{c@{\extracolsep{\fill}}cccccccc}
\toprule
$J^{P}$&\multicolumn{2}{c}{Our work}&\multicolumn{4}{c}{Refs.}\\
\cmidrule{2-3} \cmidrule{4-7}
&\multicolumn{1}{c}{A}&{B}&\cite{kepjp}&\cite{brown}&\cite{Martynenko}&\cite{Flynn}\\
\hline
$\frac{1}{2}^{+}$&	8.005&8.005	&	8.005&8.007&8.018&8.018\\
$\frac{3}{2}^{+}$&	8.049&8.049	&	8.027&8.037&8.025&8.046	\\
\bottomrule
\end{tabular*}
\end{center}

\section{Mass spectra, Regge Trajectories and Magnetic moments}

The ground state (1S), radial excited states (2S-4S) and orbital excited states (1P-5P, 1D-4D, 1F-2F) are calculated for $\Omega_{ccb}$ in this paper. We consider the total spin S=$\frac{3}{2}^{+}$ and obtained possible number of states for P, D and F are 3, 4, 4, respectively. The possible $J^{P}$ values for S states are $J^{P} =\frac{1}{2}^{+}, \frac{3}{2}^{+}$; P states are $J^{P} =\frac{1}{2}^{-},\frac{3}{2}^{-}, \frac{5}{2}^{-}$; D states are $J^{P} =\frac{1}{2}^{+}, \frac{3}{2}^{+}, \frac{5}{2}^{+}, \frac{7}{2}^{+}$; F states are $J^{P} =\frac{3}{2}^{-},\frac{5}{2}^{-}, \frac{7}{2}^{-}, \frac{9}{2}^{-}$. The masses are tabulated in Tables (\ref{tab:1}-\ref{tab:3}).  The radial and orbital excited states are determined by adding first order correction to the potential. $A$ are the masses without adding the correction and $B$ are the masses by adding first order correction. Both results are given in Tables (\ref{tab:2}-\ref{tab:3}). We do not have many theoretical prediction of masses to compare with our outcomes, though available predictions are mentioned. \\

\begin{center}
\tabcaption{\label{tab:2} Radial excited state masses of $\Omega_{ccb}$ baryon (in GeV)}
\begin{tabular*}{80mm}{c@{\extracolsep{\fill}}cccccccc}
\toprule
State&A&B&\cite{Roberts2008}\\
\hline
2S($\frac{1}{2}^{+}$)&	8.611	&	8.621&8.537	\\
3S($\frac{1}{2}^{+}$)	&		9.203	&	9.224	\\
4S($\frac{1}{2}^{+}$)	&	9.792	&	9.823	\\
5S($\frac{1}{2}^{+}$)&	10.379	&	10.424	\\
\noalign{\smallskip}\hline
2S($\frac{3}{2}^{+}$)&8.627	&	8.637&8.553	\\
3S($\frac{3}{2}^{+}$)	&	9.211	&	9.232	\\
4S($\frac{3}{2}^{+}$)&	9.797	&	9.828	\\
5S($\frac{3}{2}^{+}$)&	10.421	&	10.424	\\
\bottomrule
\end{tabular*}
\end{center} 
1S state is computed for both $J ^{P}= \frac{1}{2}^{+}$ and $J ^{P}=\frac{3}{2}^{+}$ in Table (\ref{tab:1}).  
\begin{center}
\tabcaption{\label{tab:3} Orbital excited state masses of $\Omega_{ccb}$ baryon (in GeV)}
\begin{tabular*}{80mm}{c@{\extracolsep{\fill}}cccccccc}
\toprule
State&A&B&\cite{Roberts2008}&\cite{aliev2014}&\cite{Zhi2012}\\
\hline\noalign{\smallskip}
$(1^4P_{1/2})$&		8.388	&	8.400&8.418&&8.36	\\
$(1^4P_{3/2})$&		8.372	&	8.383&8.420&8.35&8.36	\\
$(1^4P_{5/2})$&		8.354	&	8.365&8.432	\\
\noalign{\smallskip}
$(2^4P_{1/2})$&		8.970	&	8.992	\\
$(2^4P_{3/2})$&		8.954	&	8.976	\\
$(2^4P_{5/2})$&		8.934&8.955	\\
\noalign{\smallskip}
$(3^4P_{1/2})$&		9.554	&	9.585	\\
$(3^4P_{3/2})$&		9.538	&	9.569	\\
$(3^4P_{5/2})$&		9.516	&	9.547	\\
\noalign{\smallskip}
$(4^4P_{1/2})$&		10.139	&	10.181	\\
$(4^4P_{3/2})$&		10.122	&	10.164	\\
$(4^4P_{5/2})$&		10.099	&	10.140	\\
$(5^4P_{1/2})$&		10.723	&	10.775	\\
$(5^4P_{3/2})$&		10.706	&	10.758	\\
$(5^4P_{5/2})$&	10.684	&	10.735	\\
\noalign{\smallskip}
$(1^4D_{1/2})$&		8.828	&	8.848	\\
$(1^4D_{3/2})$&		8.810	&	8.831	\\
$(1^4D_{5/2})$&		8.788	&	8.808&8.568	\\
$(1^4D_{7/2})$&		8.760	&	8.780	\\
\noalign{\smallskip}
$(2^4D_{1/2})$&		9.405	&	9.437	\\
$(2^4D_{3/2})$&		9.388	&	9.420	\\
$(2^4D_{5/2})$&	9.366	&	9.396	\\
$(2^4D_{7/2})$&		9.338	&	9.368	\\
\noalign{\smallskip}
$(3^4D_{1/2})$&		9.988	&	10.031	\\
$(3^4D_{3/2})$&		9.970	&	10.012	\\
$(3^4D_{5/2})$&		9.947	&	9.988	\\
$(3^4D_{7/2})$&	9.918	&	9.957	\\
\noalign{\smallskip}
$(4^4D_{1/2})$&		10.571	&	10.622	\\
$(4^4D_{3/2})$&		10.553	&	10.605	\\
$(4^4D_{5/2})$&		10.529	&	10.582	\\
$(4^4D_{7/2})$&		10.499	&	10.553	\\
\noalign{\smallskip}
$(1^4F_{3/2})$&		9.250	&	9.280	\\
$(1^4F_{5/2})$&		9.225	&	9.254	\\
$(1^4F_{7/2})$&		9.193	&	9.222	\\
$(1^4F_{9/2})$&		9.156	&	9.184	\\
\noalign{\smallskip}
$(2^4F_{3/2})$&		9.827	&	9.864	\\
$(2^4F_{5/2})$&		9.802	&	9.840	\\
$(2^4F_{7/2})$&		9.771	&	9.809	\\
$(2^4F_{9/2})$&		9.734	&	9.773	\\
\bottomrule
\end{tabular*}
\end{center}
The obtained results are closer to the other theoretical predictions \cite{Martynenko,kepjp,Flynn,brown}. 
Refs. \cite{aliev2014} and \cite{Zhi2012} have also calculated the ground states with value 8.5 GeV and 8.23, respectively. These values are much higher than mentioned results in Table (\ref{tab:1}). \\

The orbital excited state P, D and F states are mentioned in Table (\ref{tab:3}). In 1P state, refs. \cite{aliev2014,Zhi2012} are 28 and 38 Mev less than our determined masses. 1D($\frac{5}{2}^{+}$) state value is (~200 MeV) higher than \cite{Roberts2008}. As per our knowledge, we are first to determine the F states of this triply heavy $\Omega_{ccb}$ baryon.\\

\subsection*{Regge Trajectories}

The calculated masses are used to plot the Regge trajectories for triply heavy $\Omega_{ccb}$ baryon in the $M^{2} \rightarrow$ n and $M^{2} \rightarrow J$ plane. We use
\begin{equation}
n=\beta M^2+ \beta_{0},
\end{equation}

\begin{equation}
J=\alpha M^2+ \alpha_{0},
\end{equation}

\noindent where $\beta,\alpha$, and $\beta_{0},\alpha_{0}$ are the slope and intercept, respectively, and n=\textbf{n}-1, where \textbf{n} is the principal quantum number. The values of $\beta$ and $\beta_0$ are shown in Table \ref{tab:4} and  the values of $\alpha$ and $\alpha_0$ are shown in Table \ref{tab:5}. As described in the previous section, we have calculated the masses of the S, P and D states which are used to construct Regge trajectories. The ground and radial excited states S with $J^{P}=\frac{1}{2}^{+}$ and the orbital excited state P with $J^{P}= \frac{1}{2}^{-}$, D with $J^{P}= \frac{5}{2}^{+}$ are plotted in Fig. (1) in (n, $M^{2}$) plane. The Regge trajectories for natural and unnatural parities are drawn in Fig. (2) \cite{zregge}. Straight lines were obtained by linear fitting in both figures. We observe that the square of the calculated masses fits very well to a linear trajectory and is almost parallel and equidistant in the S, P and D states. We can determine the possible quantum numbers and prescribe them to a particular Regge trajectory with the help of our obtained results.\\

\begin{figure*}
\centering
\begin{minipage}[b]{0.50\linewidth}
\includegraphics[scale=0.50]{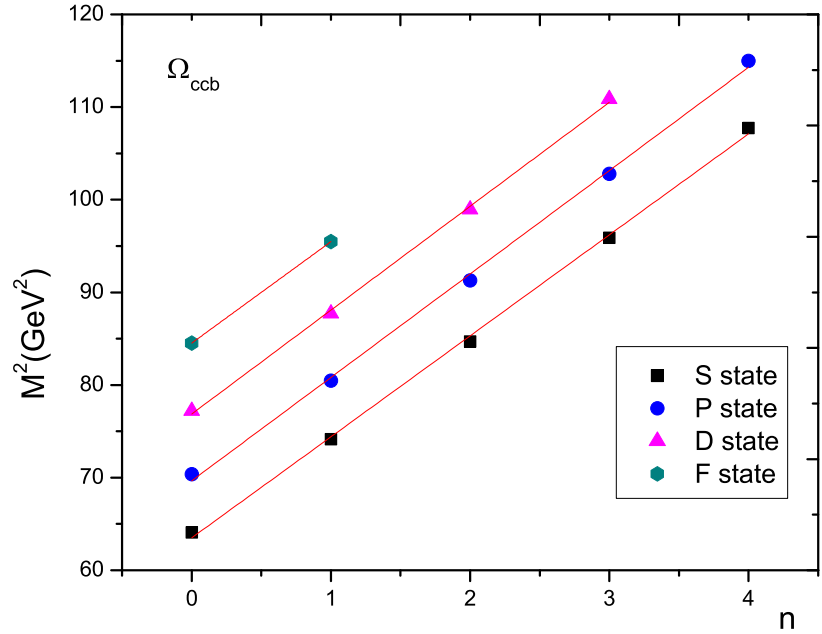}
\label{fig:1}
\end{minipage}
\caption{\label{fig:epsart} Regge trajectory in (n, $M^2$) plane.}
\end{figure*}
\begin{figure*}
\centering
\begin{minipage}[b]{0.50\linewidth}
\includegraphics[scale=0.35]{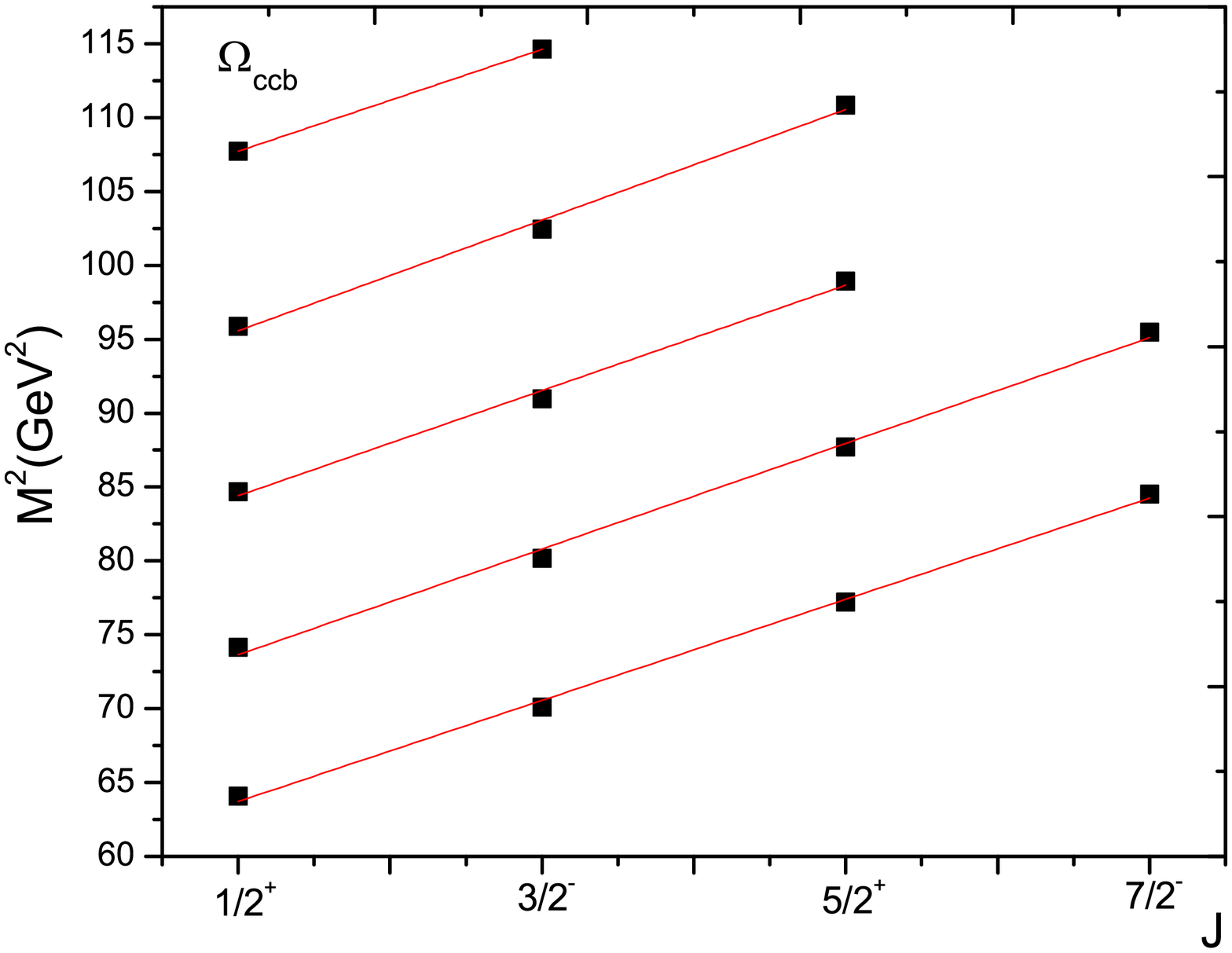}
\end{minipage}
\quad
\begin{minipage}[b]{0.40\linewidth}
\includegraphics[scale=0.35]{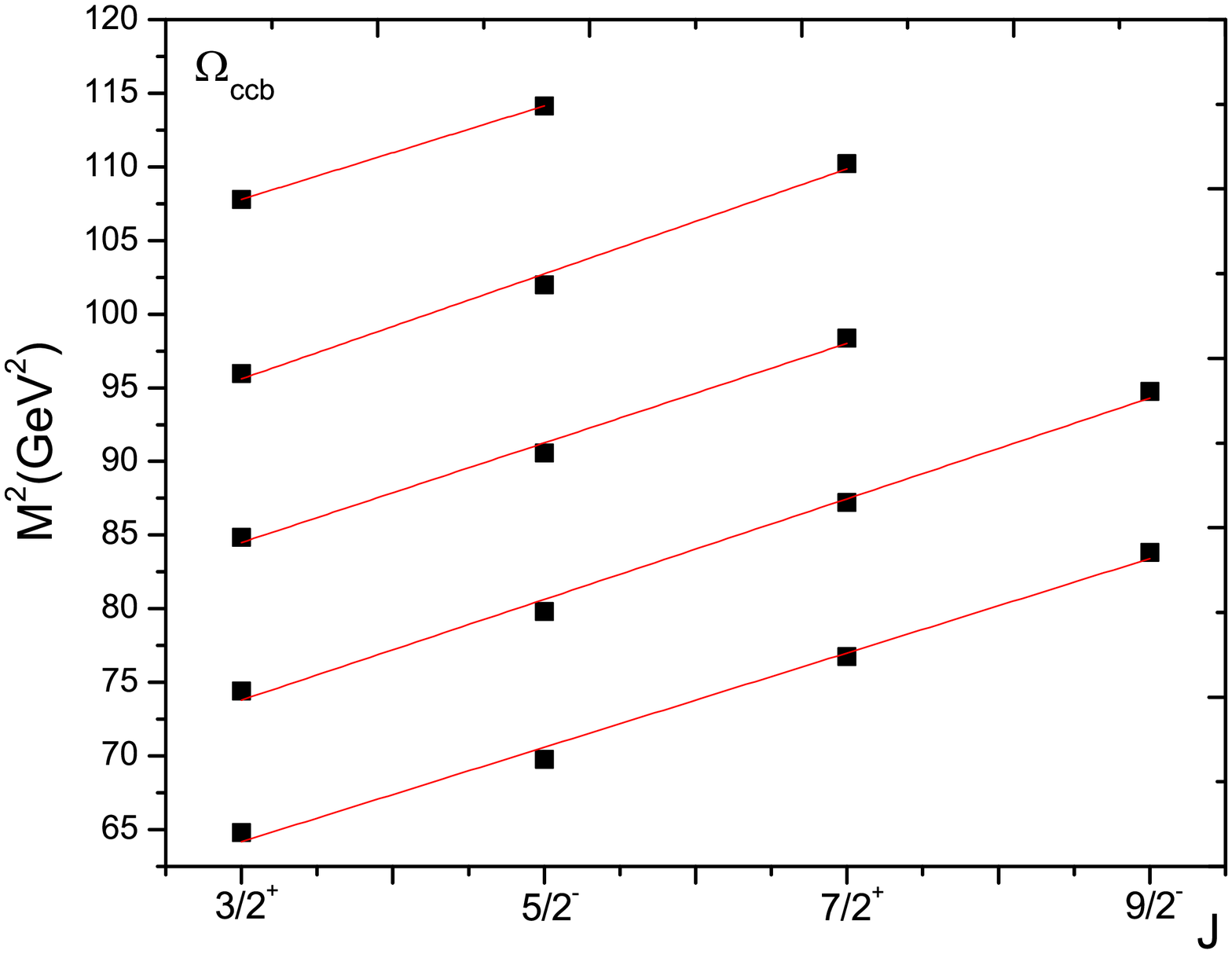}
\label{fig:2}
\end{minipage}
\caption{\label{fig:epsart} Regge trajectories (J, $M^2$) plane for natural and unnatural parities.}
\end{figure*}  
   
\begin{center}
\tabcaption{Fitted slope ($\beta$) and intercept ($\beta_{0}$) of the Regge trajectories}
\label{tab:4}
\begin{tabular*}{80mm}{c@{\extracolsep{\fill}}cccc}
\toprule
Baryon&$J^{P}$ &State& $\beta$&$\beta_{0}$\\
 \hline
 $\Omega_{ccb}$&$\frac{1}{2}^{+}$&S&0.092$\pm$0.0017&-5.816$\pm$0.151\\
&$\frac{1}{2}^{-}$&P&0.0894$\pm$0.0019 &-6.231$\pm$ 0.177\\
&$\frac{5}{2}^{+}$&D&0.089$\pm$0.0017&-6.848$\pm$0.169\\
\hline
\end{tabular*}
\end{center}

\begin{center}
\tabcaption{Fitted slope ($\alpha$) and intercept ($\alpha_{0}$) of the Regge trajectories}
\label{tab:5}
\begin{tabular*}{80mm}{c@{\extracolsep{\fill}}cccc}
\toprule
Baryon&$J^{P}$ &State& $\alpha$&$\alpha_{0}$\\
 \hline
 $\Omega_{ccb}$&$\frac{3}{2}^{+}$&S&0.090$\pm$0.002&-4.725$\pm$0.191\\
&$\frac{5}{2}^{-}$&P&0.090$\pm$0.002 &-5.218$\pm$ 0.181\\
&$\frac{7}{2}^{+}$&D&0.089$\pm$0.0017&-5.835$\pm$0.166\\
\hline
\end{tabular*}
\end{center}

\subsection*{Magnetic moments }

\begin{table*}
\begin{center}
\tabcaption{\label{tab:6} Magnetic moments (in nuclear magnetons) }
\begin{tabular*}{100mm}{c@{\extracolsep{\fill}}ccccccc}
\toprule
Baryons& function\cite{107}&Our&\cite{103}&\cite{101}&\cite{bhavin}&\cite{kepjp}\\
\hline
$\Omega_{ccb}^{+}$&$\frac{4}{3} \mu_{c}$- $\frac{1}{3} \mu_{b}$&0.606&0.505&0.522&0.502&\\
\hline
$\Omega_{ccb}^{* +}$&$\mu_{b}$+2$\mu_{c}$&0.819&0.659&0.703&&0.807\\
 \bottomrule
\end{tabular*}
\end{center}
\end{table*}

The magnetic moments of baryons are obtained in terms of the spin, charge and effective mass of the bound quarks as \cite{bhavin,zcpc}
\begin{eqnarray}\nonumber
\mu_{B}=\sum_{i}\langle \phi_{sf}\vert \mu_{iz}\vert\phi_{sf}\rangle)
\end{eqnarray}
where
\begin{equation}
\mu_{i}=\frac{e_i \sigma_i}{2m_{i}^{eff}},
\end{equation}
where $e_i$ is a charge and $\sigma_i$ is the spin of the respective constituent quark corresponding to the spin flavor wave-function of the baryonic state.  The effective mass for each of the constituting quarks $m_{i}^{eff}$ can be defined as
\begin{equation}
m_{i}^{eff}= m_i\left( 1+ \frac{\langle H \rangle}{\sum_{i} m_i} \right)
\end{equation}
where $\langle H \rangle$ = E + $\langle V_{spin} \rangle$. 
The spin flavor wave function and magnetic moments of the baryon is given in Table ({\ref{tab:6}}).

\section{Conclusion}
We discuss the baryon with the combination of two charm quark and one beauty quark in this paper. The $\Omega_{ccb}$  baryon is unknown from experimental side. The mass spectra of $\Omega_{ccb}$ baryon are listed using the Hypercentral Constituent Quark Model starting from S state to F state. We also added the first order correction to the potential (see Eqn. (6)), the results without correction(A) and with correction(B) are tabulated in Tables (\ref{tab:1}-\ref{tab:3}).  The ground states have already been studied by various models and the results of some of them are in same range of our prediction. Although, the excited states are calculated by very few theorist. We also notice that first radial(2S) and orbital(1P) excited states masses are in range with other predictions. While for higher excited state 1D, the difference is higher. The 2S state is only determined by Roberts et al. \cite{Roberts2008} as per our knowledge.  Their values are 74(84) MeV lesser for 2S state. Although, the difference between the isospin doublet states are 16 MeV in \cite{Roberts2008} and in similar manner our 2S state masses ($\frac{3}{2}^{+}$- $\frac{1}{2}^{+}$) also have 16 MeV difference. Thus, the obtained masses are very much useful to obtain the resonances of higher excited states.  \\

The first order correction for the excited state varies from 0.1\% - 0.4\% for 1S-4S and 1P-4P states; 0.2\% - 0.5\% for 1D-4D states; 0.3\% for F states. It is clear from the results that as we move towards the higher excited states, the contribution of the correction is increasing simultaneously. It will be interesting to see this effect of correction in the light sector  baryons \cite{aip}.\\

The paper also features the values of ground state magnetic moments of $\Omega_{ccb}$ and $\Omega_{ccb}^*$ baryons. Our obtained results are reasonably close to the other predictions. The acquired Regge trajectories will also be helpful to define unknown states of $\Omega_{ccb}$ baryon and $J^{P}$ values can be assigned.

\vspace{1mm}

\end{multicols}

\vspace{-1mm}
\centerline{\rule{80mm}{0.1pt}}
\vspace{2mm}

\begin{multicols}{2}

\end{multicols}

\clearpage

\end{document}